\documentstyle[prl,aps]{revtex}
\begin{document}
\draft

\begin{twocolumn}
\title{Absence of hysteresis in the heat capacity of the three-dimensional
random-field Ising model}
\author{J. Satooka\thanks{present address: The Institute of Physical and Chemical Research (RIKEN), 
Wako, Saitama 351-0198, Japan }}
\address{The Doctoral Research Course in Human Culture, Ochanomizu
University, Bunkyo-ku, Tokyo 112-8610, Japan}
\author{H. Aruga Katori, A. Tobo and K. Katsumata}
\address{The Institute of Physical and Chemical Research (RIKEN), 
Wako, Saitama 351-0198, Japan}
\date{\today}
\maketitle

\begin{abstract}
Heat capacity measurements on the prototypical three-dimensional random-field Ising model compound Fe$_{0.58}$Zn$_{0.42}$F$_{2}$ have been performed in both of the field-cooling (FC) and zero-field-cooling (ZFC) procedures. 
There is no evidence of hysteresis between the FC and ZFC protocols within the experimental accuracy.
The absence of hysteresis is interpreted based on a magnetic field $(H)-$temperature $(T)$ phase diagram in which a metastability region exists between the N\'eel temperature line $T_{N}(H)$ and $H_{c}^{(1)}(T)$, the latter being the lowest critical field line for the local spin flips.  
\end{abstract}
 
The random-field Ising model (RFIM) has been a subject of extensive studies over the last two decades.
It has been predicted theoretically\cite{F79} that the RFIM can be realized in diluted uniaxial antiferromagnets under an applied magnetic field.
Since then, a number of experimental studies has been done on this topic.
Among them, the three-dimensional (3D) RFIM has attracted much attention.
The 3D RFIM shows distinctly different behavior depending on whether the measurements are done with cooling the sample in a
magnetic field (field-cooling, FC) or done in a magnetic field with increasing temperature after having cooled it in zero field (zero-field-cooling, ZFC) below the transition temperature.
Early neutron scattering experiments on the 3D RFIM showed that a long-range magnetic ordering was established in the ZFC case,
whereas the system was in a disordered domain state in the FC
case\cite{BKJ85,YC85}.

It turns out that the phase transition of the 3D RFIM is more complex.
Jaccarino $et~al$.\cite{JKB85} have proposed a temperature ($T$) versus magnetic field ($H$) phase diagram of the 3D RFIM based on experimental facts.
They showed that a metastability region exists in the $H-T$ plane
bordering on the second-order phase transition temperature line $T_{N}(H)$.
From a synchrotron magnetic X-ray scattering study of the 3D RFIM, Hill $et~al$.\cite{HT91,HF93,H97} showed that the long-range-order (LRO) established in ZFC procedure vanishes continuously with increasing temperature in the vicinity of $T_{N}(H)$.
They labeled this as "trompe l'oeil" critical behavior.
There has been a big debate\cite{W2338,HFB96,W2340,BKM96,BFH96} on the nature of the phase transition of the 3D RFIM in a recent issue of this Journal. So, the topic may be interesting to broad audiences.

The phase transition in the prototypical examples of the 3D RFIM
Fe$_{x}$Zn$_{1-x}$F$_{2}$ has been studied by indirect and direct heat capacity measurements\cite{BKJ82,DB89,FK91}.
From optical linear birefringence ($\Delta n$) measurements, Belanger $et~al$.\cite{BKJ82} have observed a sharp peak in the temperature dependence of $d(\Delta n)/dT$ of a Fe$_{0.53}$Zn$_{0.47}$F$_{2}$ sample in finite magnetic fields ($H\leq 2~\rm T$).
Later, Dow and Belanger\cite{DB89} have measured the heat capacity ($C_{p}$) of a Fe$_{0.46}$Zn$_{0.54}$F$_{2}$ crystal using a variation of the classical heat pulse technique in $H$ up to 1.5 T. They found an apparent rounding of the peak in $C_{p}$ for the FC measurement.
This rounding of $C_{p}$ for the FC procedure has also been observed in the $\Delta n$ study\cite{FK91} up to 1.9 T.
Birgeneau $et~al$.\cite{BF95} have measured $C_{p}$ of a Fe$_{0.5}$Zn$_{0.5}$F$_{2}$ crystal using a standard semiadiabatic dc calorimeter.
They found no difference in $C_{p}$ for FC and ZFC protocols in $H$=1.5 and 5.5 T.
We have made a comprehensive study of $C_{p}$, using a relaxation method, to clarify this apparently contradictory situation.
In this paper, we report the results obtained in a single crystal of Fe$_{0.58}$Zn$_{0.42}$F$_{2}$.

The non-diluted compound FeF$_{2}$ has the rutile-type crystal structure $D^{14}_{4h}-P4/mnm$~\cite{S54} and it establishes an antiferromagnetic LRO below the N\'eel temperature $T_{N}$=78.4 K~\cite{S55}.
The spin easy axis is parallel to the $c$ axis. A large single ion anisotropy makes this an excellent example of a 3D Ising system.
The single crystal of Fe$_{0.58}$Zn$_{0.42}$F$_{2}$ used in this study had been grown at the University of California, Santa Barbara.

The heat capacity was measured using a MagLab$^{\rm HC}$ microcalorimeter of Oxford Instruments, UK.
This microcalorimeter consists of a small sapphire chip on which a serpentine metallic heater is evaporated.
Attached to the chip with 50 $\mu $m gold leads is a tiny temperature sensor.
The chip is suspended by 20 $\mu $m tungsten leads which make electrical connections to the elements and also provide a weak thermal link to a calorimeter cell.
The cell in turn is screwed onto the $^{3}$He pot with thermally conductive grease.
For a given cell temperature, the chip temperature is a function
(only) of the power dissipated in the heater element.
For a measurement, the heater power was first increased stepwise (typically $\sim $10 $\mu $W near the N\'eel temperature) and maintained for a period of 3$\tau $ to ensure equilibrium was closely approached, where $\tau $ is the relaxation time
(typically $\sim $70 sec near the N\'eel temperature).
Then the power was stepped back to its original value. 
The variation of the chip temperature with time was fitted with an exponential function from which $\tau $ was obtained.
We repeated this procedure several times to improve the quality
of the data.
With this calorimeter, FC and ZFC measurements can be done with the same accuracy.
The single crystal was cut into a platelet parallel to the $c$ axis with the dimensions about 1.5 mm${\times}$3 mm${\times}$0.1 mm.
The platelet was mounted on the sapphire chip using a small amount of thermally conductive grease.
The heat capacity of the sample was obtained by subtracting the heat capacity of the sapphire chip from the total.
The heat capacity of the sapphire chip is much smaller ($\sim $1/10) than that of the sample in the temperature range of interest.
In both of the ZFC and FC measurements, the sample was cooled from $T$=90 K which is higher enough than $T_{N}$ in pure FeF$_{2}$.

We measured the temperature dependence of $C_{p}$ of
Fe$_{0.58}$Zn$_{0.42}$F$_{2}$ in $H$ between 0 T and 10 T parallel to the $c$ axis.
Typical results under ZFC condition are shown in Figs.\ref{fig1}
(a) and (b).
Here, we subtracted the contribution of the lattice from the
total using $C_{p}$ of ZnF$_{2}$\cite{S55}.
The result after subtraction corresponds to the magnetic part ($C_{mag}$) of the heat capacity.@
In zero field (ZF) an asymmetric peak appears at 45.5 K.
We define the peak temperature, $T_{p}(H)$, as the temperature at which $C_{mag}$ is a maximum.
As is seen from Fig.\ref{fig1}, $T_{p}(H)$ decreases and the shape
of $C_{mag}$ becomes symmetric and broad with increasing $H$.
These observations are consistent with the results measured on the
Fe$_{0.53}$Zn$_{0.47}$F$_{2}$ ($H\leq 2~\rm T$)\cite{BKJ82},
Fe$_{0.46}$Zn$_{0.54}$F$_{2}$ ($H\leq 1.9~\rm T$)\cite{FK91} and
Fe$_{0.6}$Zn$_{0.4}$F$_{2}$ ($H\leq 8~\rm T$)\cite{FK91} samples with the $\Delta n$ method under ZFC condition.

Figure \ref{fig2} shows the data taken at $H$=2 T for both of the FC and ZFC protocols. It is evident that there is no difference in the two
measurements within the experimental accuracy ($\sim $0.3 K).
We have observed no hysteresis in other fields investigated (0.5, 1.0, and 3.0 T).
This result is in accord with that reported by Birgeneau $et~al$.\cite{BF95}.
However, the present result is in contradiction with the observation by Dow and Belanger\cite{DB89} and by Ferreira $et~al$.\cite{FK91}, in which FC heat capacity is severely rounded.

These apparently contradictory results may be explained as a result of the time scale of the measurements.
It is widely accepted that the critical dynamics of the 3D RFIM is quite slow.
Then, the results obtained with different technique such as $\Delta n$, adiabatic or relaxation method may be different.
The time scale of our measurement is typically 200 sec.
Moreover, we can estimate the equilibrium value by fitting the time evolution with an exponential function.

Belanger $et~al$. have pointed out that the concentration gradient of the sample makes the observation of hysteresis difficult~\cite{BK88}. 
In case of the measurement using a sample the large concentration gradient, it was reported that the peak of $C_{mag}$ in ZF was rounded. 
On the contrary, the peak of our heat capacity data measured in ZF is sharp enough to suggest that the concentration gradient of our sample is very small. 
In order to confirm this, we analyzed the data measured with ZF by fitting them to the scaling function~\cite{BC83}
\begin{equation}
 C_{mag}=A/\alpha \mid \tau \mid ^{-\alpha }(1+D\mid \tau \mid ^{X})+B,
\label{eqn:1}
\end{equation}
where $\tau $=$T$/$T_{N}$-1. 
The fitting parameters thus obtained close to those reported from the analysis of the indirect heat capacity data, 
$d(\Delta n)/dT$, measured in ZF using the sample of Fe$_{0.60}$Zn$_{0.40}$F$_{2}$~\cite{BC83}. 
Thus, we are convinced that our sample is homogeneous enough to observe the random field effects.

Next, we discuss the apparent rounding of the peak at high fields
(Fig.\ref{fig1} (b)).
In Fig.\ref{fig3}, we show the field dependence of the width ($W_{p}$) of the peak of $C_{mag}$ defined as the one at which
the value of $C_{mag}$ is 95 \% of its maximum in respective $H$ obtained from the ZFC procedure.
As is seen from Fig.\ref{fig3}, $W_{p}$ is almost constant below 4 T.
On the other hand, $W_{p}$ depends much on $H$ above 4 T.
We interpret this broadening of the peak of $C_{mag}$ as an evidence that the system is in a domain state at finite fields.

King $et~al$.\cite{KJ81} have observed in Fe$_{x}$Zn$_{1-x}$F$_{2}$ that the magnetization shows anomalies at $H_{c}^{(n)}=(n/8)H_{E}$, where $H_{E}$ is the exchange field and $n$=1, 2, ... 5.
These anomalies have been explained as arising from local spin flips which occur when the external magnetic field becomes equal to the local exchange fields $H_{c}^{(n)}$.
The lowest field $H_{c}^{(1)}$ required for the flip is about 7 T at 1.3 K\cite{KJ81}.
Since the exchange field is proportional to the thermal average of the magnetic moment, it decreases with increasing temperature and vanishes at $T_{N}$.
When the sample is cooled in zero field below $T_{N}$ and $H=H_{0}$ is
subsequently applied, the system is in the LRO state provided
$H_{0}<H_{c}^{(1)}(T)$.
At an elevated temperature, $H_{c}^{(1)}(T)$ becomes equal to $H_{0}$, where the local spin flip occurs.
This spin flip will nucleate domains. Near the N\'eel temperature, $H_{c}^{(n)}(T)$ with $n$=1, 2, ... become close, thereby domains with various sizes will be nucleated.
It is not surprising that the "trompe l'oeil" critical behavior has been observed in the 3D RFIM at finite fields\cite{HF93,BF95}.

When the sample is cooled from high temperature under a magnetic field (FC case) and $T_{N}(H)$ is reached, a LRO will be
established there.
However, as is discussed above, $H_{c}^{(n)}(T)$ with $n$=1, 2, ... are close near $T_{N}(H)$, thence domains will be nucleated immediately
below $T_{N}(H)$.
These domains are stable at low temperatures against thermal agitation because the anisotropy is strong.
According to this interpretation, the transition temperature in equilibrium is the N\'eel temperature $T_{N}(H)$ and there is a metastability region below $T_{N}(H)$ bounded by the $H_{c}^{(1)}(T)$ line. 
This conclusion is consistent with the result of the computer calculation\cite{SG85} made on a diluted 3D Ising antiferromagnet in magnetic fields.

Because the system is in the domain state at this metastability region for both of the ZFC and FC conditions, the heat capacities coming from thermal fluctuation in these states are expected to be similar.
This explains the absence of hysteresis in the ZFC and FC heat capacities as observed in the present experiment.
Based on this interpretation, then, it is possible to observe the crossover from the random exchange Ising model to RFIM with
increasing $H$ along the $T_{N}(H)$ line.

In the remaining part of this paper, we discuss the shift in the transition temperature with $H$.
Here, we assign the peak temperature $T_{p}(H)$ as the transition temperature.
The true transition temperature $T_{N}(H)$ may be a bit larger than $T_{p}(H)$ for a given $H$.
However, as is seen from Fig.\ref{fig3}, the difference between $T_{p}(H)$ and $T_{N}(H)$ lies within the experimental error below 4 T.

In Fig.\ref{fig4}, $\Delta T_{p}(H)(=T_{p}(H=0)-T_{p}(H))$ is plotted as a function of $H$ on a log-log scale.
Data were fitted to Eq.(2) below over the field range $0.5~\rm T\leq $$H$$\leq 4.0~\rm T$ from the reason discussed above,
\begin{equation}
 \Delta T_{p}(H)=aH^{2/\phi }+bH^{2},
\label{eqn:2}
\end{equation}
where $a$ and $b$ are constants and $\phi $ is the crossover exponent.
The term $bH^{2}$ represents a small mean-field shift. 
From the fitting, we got $\phi =1.39\pm 0.03$.
This value is close to the ones observed experimentally\cite{BKJ82,FK91,YM82} and the theoretical
ones ($\phi $=1.25\cite{F79}$\sim $1.4\cite{A86}).

In conclusion, we have made a comprehensive study of $C_{p}$, using a relaxation method, on the prototypical 3D RFIM compound
Fe$_{0.58}$Zn$_{0.42}$F$_{2}$ in both of the FC and ZFC procedures.
We have observed no hysteresis in the two measurements within the experimental accuracy.
We interpret this result based on an $H-T$ phase diagram in which
a metastability region exists between $T_{N}(H)$ and $H_{c}^{(1)}(T)$, the latter being the lowest critical field for the local spin flip.
This phase diagram seems to settle some of the conflicting experimental results reported so far.

We would like to thank A. Aharony, D. P. Belanger, R. J. Birgeneau A. Ito and W. Kleemann for helpful discussions. J. S. would like to thank A. Ito for her encouragements.

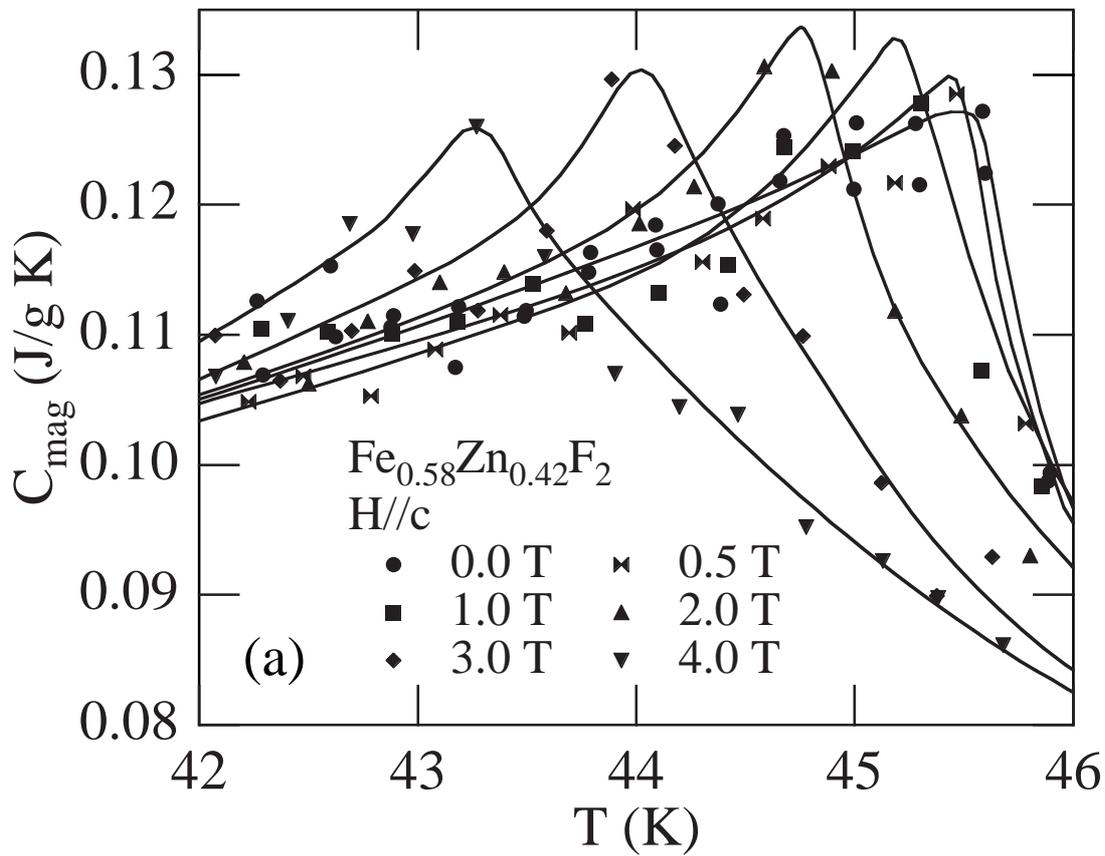
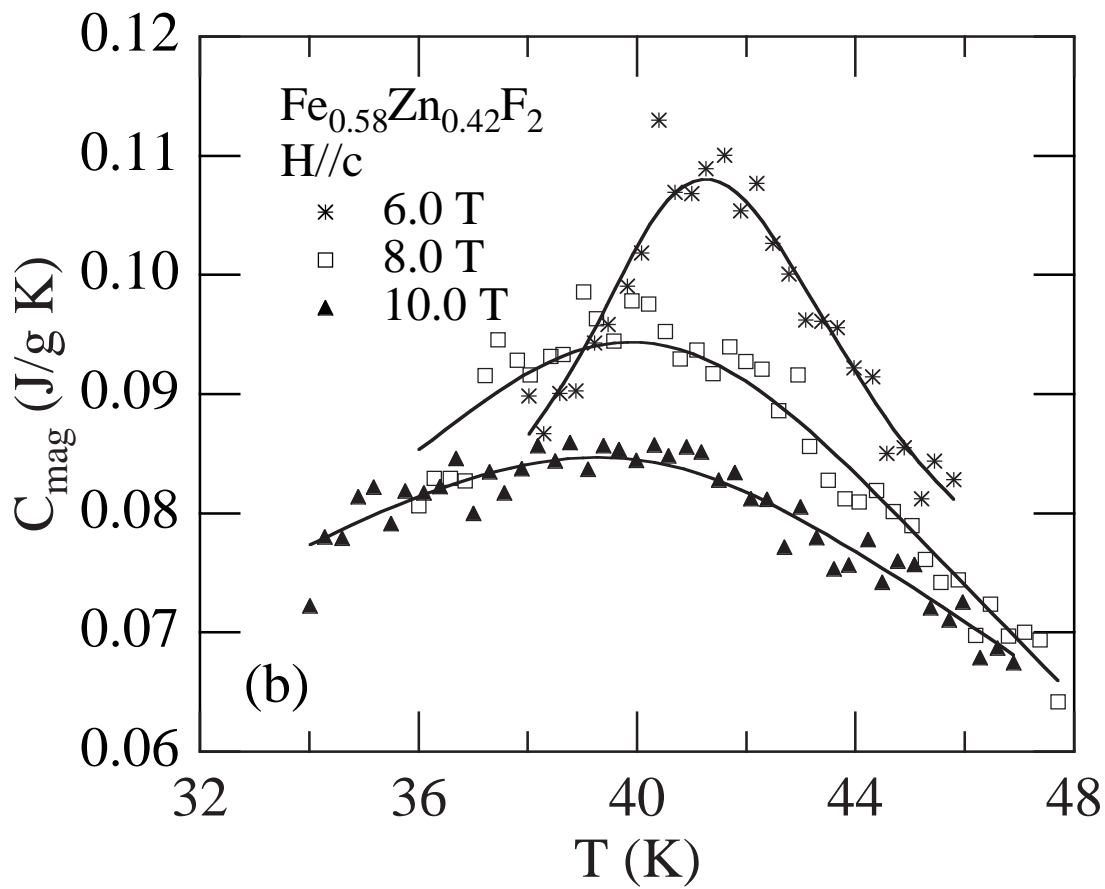
\begin{figure}
\caption{Temperature dependence of the magnetic heat capacity in a
Fe$_{0.58}$Zn$_{0.42}$F$_{2}$ single crystal obtained under ZFC condition. The external magnetic field is applied parallel to the $c$ axis.}
\label{fig1}
\end{figure}

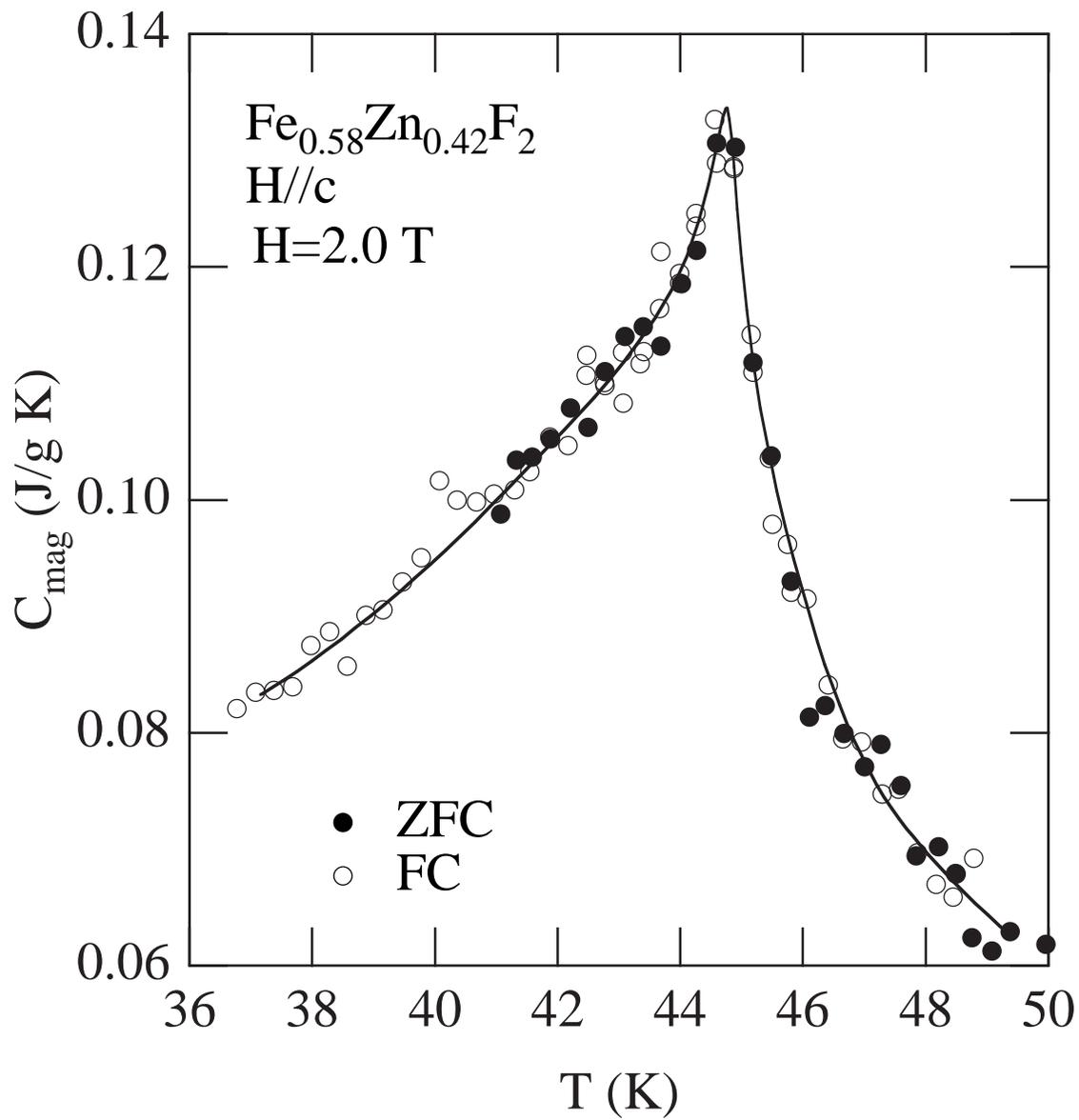
\begin{figure}
\caption{Temperature dependence of the magnetic heat capacity in a
Fe$_{0.58}$Zn$_{0.42}$F$_{2}$ single crystal taken at 2 T for FC and ZFC protocols.}
\label{fig2}
\end{figure}

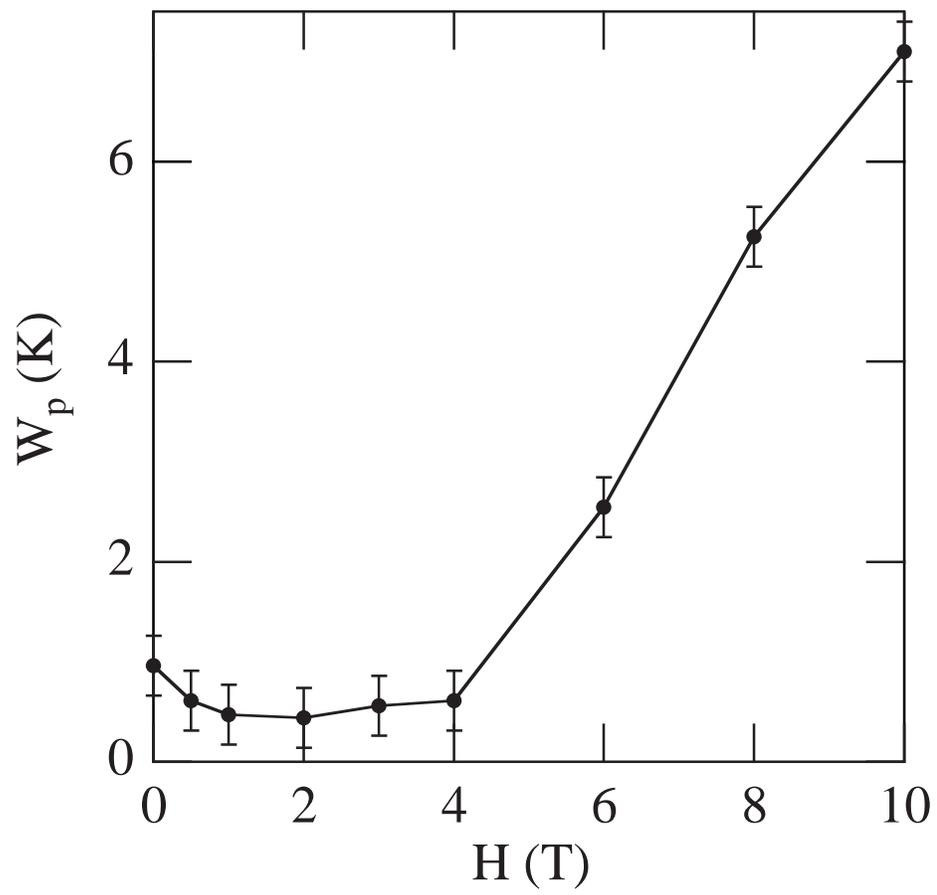
\begin{figure}
\caption{Field dependence of the width of the peak in the magnetic heat capacity of a Fe$_{0.58}$Zn$_{0.42}$F$_{2}$ single crystal. }
\label{fig3}
\end{figure}

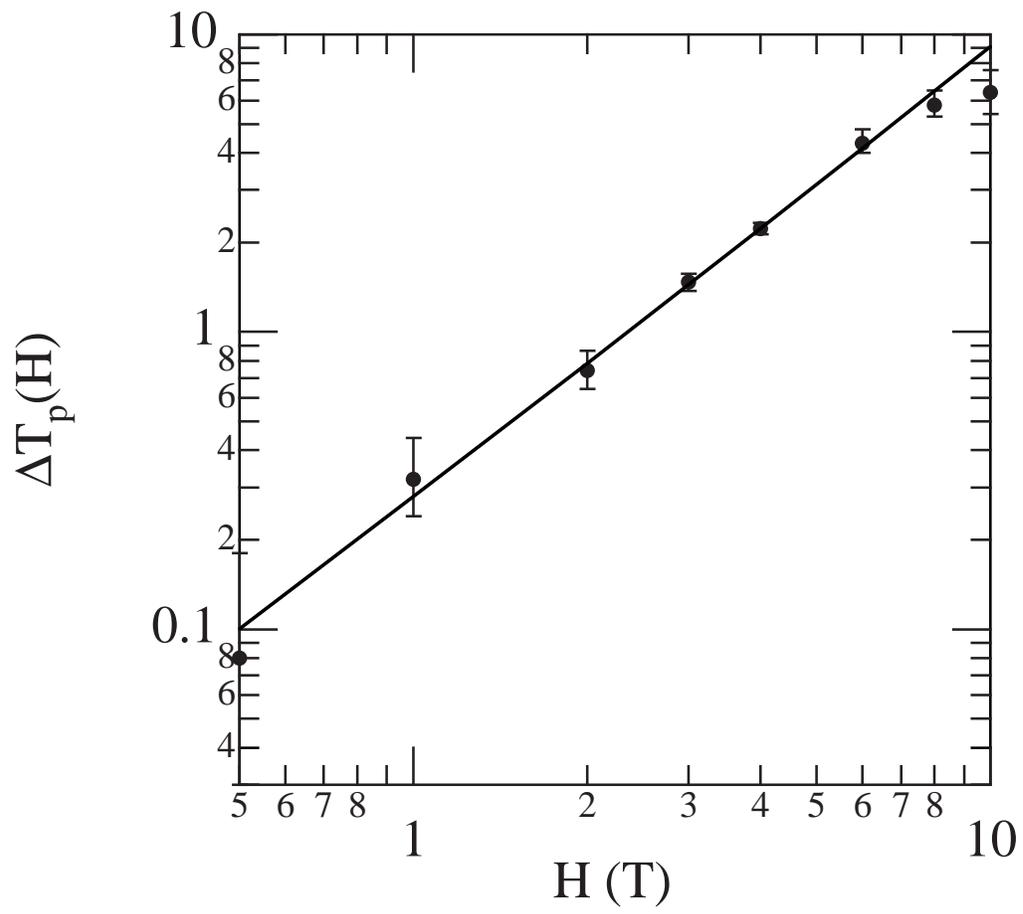
\begin{figure}
\caption{Shift of the peak temperature in the magnetic heat capacity of a Fe$_{0.58}$Zn$_{0.42}$F$_{2}$ single crystal with magnetic field.}
\label{fig4}
\end{figure}

\end{twocolumn}
\end{document}